\documentstyle[twoside,twocolumn,prl,floats,aps]{revtex}
\tighten 
\begin{document}
\draft

\title{Quantum dot to disordered wire crossover: A complete solution \\
in all length scales for systems with unitary symmetry}

\author{A. M. S. Mac\^edo}

\address{ 
Laborat\'orio de F\'{\i}sica Te\'orica e Computacional, Departamento de
F\'{\i}sica,\\  Universidade Federal de Pernambuco\\
50670-901 Recife, PE, Brazil\\
{\rm (Submitted 29 June 1999)}\\
\parbox{14 cm}{\medskip\rm\indent
We present an exact solution of a supersymmetric non-linear $\sigma$-model
describing the crossover between a quantum dot and a disordered  quantum wire
with unitary symmetry.  The system is coupled ideally to two electron
reservoirs via perfectly conducting leads sustaining an arbitrary number of
propagating channels.  We obtain closed expressions for the first three
moments of the conductance, the  average shot-noise power and the
average density of transmission eigenvalues. The results are complete in the
sense that they are nonperturbative and are valid in all regimes and length
scales. We recover several known results of the recent literature by taking
particular limits. }}
\maketitle
Transport in phase coherent devices has become a rather topical subject, ever 
since modern nanolithographic techniques have achieved sufficient precision
to manufacture a great variety of such systems \cite{fg97}. These mesoscopic
structures can be broadly divided into two categories: (I) disordered
conductors, where static defects generates disorder scattering with an elastic
mean free path, $l$, smaller then the device dimensions and (II) ballistic
cavities, where $l$ exceeds the device dimensions and boundary scattering
dominates. The hallmark of mesoscopic phenomena is the relevance of quantum
interference effects due to multiply scattered waves. These in turn implies
that the only really important feature of the microscopic scattering mechanism
is the preservation of phase memory.  
\par  
In spite of the
many similarities of mesoscopic effects in systems of type I and II, the
microscopic theoretical models to describe them are quite different. In the
most understood regime (the semiclassical), impurity average diagrammatic
techniques \cite{lsf87} are used to study disordered conductors, whilst quantum
chaotic scattering theory and trace formulas are used to describe transport in
ballistic cavities \cite{bjs93}. There are many alternative (non-microscopic)
approaches to quantum transport that substitute the  detailed information on
the scattering potentials by more generic statistical hypothesis on macroscopic
quantities, e.g. by assuming independent randomness in the scattering matrix
of a macroscopic portion of the system.  
\par   
These alternative theories (known generically as random-matrix models
\cite{gmh97,b97}) have proved exact in many situations and are very useful in
describing non-perturbative mesoscopic effects. The most remarkable aspect of
these random-matrix models  is the existence of an underlying mathematical
structure that captures the relevant statistical information to the
description of mesoscopic phenomena, which is known as the supersymmetric
non-linear $\sigma$-model  and has been introduced in condensed matter physics
through the pioneering work by Efetov \cite{e83}. For disordered
systems non-linear $\sigma$-models of various kinds have been rigorously
derived from microscopic models and succesfully applied to the
non-perturbative regime \cite{e83,e97}. More recently, derivations of
non-linear $\sigma$-models for ballistic chaotic cavities have been
presented \cite{mk95,aasa96}. The crucial underlying concept of these
recent works is the existence of an intimate connection between the
irreversible classical dynamics and the statistical quantum properties of the
system. 
\par 
The inevitable conclusion is that the non-linear $\sigma$-model is a key
concept in the search for a unified description of systems of type I and II.
Earlier atempts in this direction have been confined to the semiclassical
regime \cite{iwz90} or to the thick-wire limit \cite{z92}. In this letter
we provide the complete picture by incorporating all physical regimes
and useful limits. Specifically, we use a supersymmetric non-linear
$\sigma$-model  to describe the continuous crossover between a quantum dot
(ballistic chaotic cavity) and a quantum wire (quasi-one-dimensional disordered
conductor) in the absence of time-reversal symmetry. The system is coupled
ideally (no transmission barriers) to two electron reservoirs through
perfectly conducting leads that support and arbitrary number of open channels.
We find an exact solution of the model and obtain closed expressions for
several important transport properties. We show that our solution reproduces
many known results of the literature by taking particular limits.    
\par
Our starting point is the
Landauer-B\"uttiker formalism, by which  transport observables can be related
to scattering functions, such as the  transmission matrix $t$ of the system. 
Let $N_1$ and $N_2$ denote the number of channels in lead 1 and 2 respectively.
We define $t$ as an $N_1 \times N_2$ random transmission matrix describing the
ideal coupling of the modes in lead 1 to those in lead 2 though a
disordered sample of length $L$. In the limit $L \to 0$ the system
represents a quantum dot or a ballistic chaotic cavity, in which disorder
scattering is negligible and the irregular fluctuations in the transmission
matrix are due to surface scattering at the boundaries of the cavity. We shall
measure the sample's size $L$ in units of the localization length $\xi$, which
we take as a free parameter, so that $s\equiv 2L/\xi$ is a dimensionless
variable controling the crossover from the ballistic ($s=0$) to the insulating
regime ($s \gg 1$). The mean free path, $l$, sets the typical size of the
cavity, so that when the disordered portion has length $L>0$ the combined
dot-wire system has length $L_{sys} > l$.
For $\xi \gg l$ (large number of scattering channels) the system exhibits a
metallic (or diffusive) behaviour which interpolates between the ballistic and
the insulating regimes. 
\par
A conventional approach to quantum transport is to study the statistical
properties of a particular observable, such as the system's conductance.
Here, we shall instead consider a generating function \cite{n94,r96,bf96} from
which many transport properties can be derived (such as the first three
moments of the conductance or the average shot-noise power). Consider thus the
following function  
\begin{equation} 
{\cal Z}(\{\theta \})={\rm det}\left( \frac{1-\sin ^2(\theta 
_0/2)tt^{\dagger }}{1+\sinh ^2(\theta _1/2)tt^{\dagger }}\right) 
\label{genfun}  
\end{equation} 
and let $\Psi_s (\{\theta \})$ be
the average of  ${\cal Z}(\{\theta \})$ over realizations on the ensemble
of random transmission matrices. Note that the joint distribution of the
eigenvalues of $tt^{\dagger}$ contains $s$ as a parameter. Using a non-linear 
$\sigma$-model to be presented later we shall prove that  
\begin{eqnarray} 
\Psi_s (\{\theta \}) &=&1+(\cos \theta _0-\cosh \theta 
_1)\sum_{n=0}^{N-1}\int_0^\infty d\mu _{nk}c_{nk}^{(N)}c_{nk}^{(M)} 
\nonumber 
\\ 
&&\times P_n(\cos \theta _0) P_{-1/2+ik}(\cosh \theta _1)e^{-\varepsilon 
_{nk}s},  
\label{central} 
\end{eqnarray} 
where $d\mu _{nk}=(\varepsilon _{nk})^{-1}(2n+1)k\tanh \left( k\pi \right)dk$ 
is the integration measure, $\varepsilon _{nk}= k^2+1/4+n(n+1)$, $N={\rm
min}(N_1,N_2)$, $M={\rm max}(N_1,N_2)$, $P_n(x)$ is the Legendre
polynomial, $P_{-1/2+ik}(x)$ is the conical function and     
\[ 
c_{nk}^{(N)}=\frac{\left| \Gamma (N+1/2+ik)\right| ^2}{(N-n-1)!(N+n)!}. 
\] 
Equation (\ref{central}) is the central result of this letter. 
It constitutes the exact solution of a realistic model of quantum
transport that applies to all regimes of physical interest, namely:
ballistic, diffusive, insulating, semiclassical and the extreme quantum limit.
Furthermore, it contains the exact solution of important
models, such as the  thick wire \cite{z92,mgz94} and the DMPK theory 
\cite{dmpk}, as particular cases.
In order to 
demonstrate the usefulness of our explicit solution, we shall now discuss some
concrete applications.   
\noindent 
(i) Moments of the conductance 
 
According to the Landauer-B\"utikker scattering approach, the two probe 
conductance of a phase coherent system is given by  $G=G_0{\rm tr}(tt^{\dagger
})$, where $G_0=2e^2/h$ is the conductance quantum. It is easy to verify
from (\ref{genfun}) that the  first three moments of the conductance can be
obtained from $\Psi_s (\{\theta \})$ through the relations: 
\[ 
\langle G/G_0\rangle _s=2\left( D_{\theta _0}\Psi_s (\{\theta 
\})\right)\Big| _{\{\theta \}=0} 
\] 
\[ 
\langle \left( G/G_0\right) ^2\rangle _s=-4\left( D_{\theta 
_0}D_{\theta _1}\Psi_s (\{\theta \})\right)\Big| _{\{\theta \}=0} 
\] 
\[ 
\langle \left( G/G_0\right) ^3\rangle _s=4\left( D_{\theta 
_0}D_{\theta _1}(D_{\theta _1}-D_{\theta _0})\Psi_s (\{\theta \})\right)\Big| 
_{\{\theta \}=0} 
\] 
where $D_\theta \equiv \partial/\partial (\cos \theta )$. Using Eq. (\ref{central}) 
we obtain the compact result 
\begin{equation} 
\left\langle \left( G/G_0\right) ^m\right\rangle 
_s=2\sum_{n=0}^{N-1}\int_0^\infty d\mu 
_{nk}c_{nk}^{(N)}c_{nk}^{(M)}g_{nk}^{(m)}e^{-\varepsilon _{nk}s} ,
\label{moments} 
\end{equation} 
where $g_{nk}^{(1)}=1$, $g_{nk}^{(2)}=k^2+1/4+n(n+1)$ and 
$g_{nk}^{(3)}=\frac 14(k^2+1/4)(k^2+9/4)+2n(n+1)(k^2+1/4)+\frac 14%
(n+2)(n+1)n(n-1)$. 
\par
This expression reproduces (for $m=1,2$) the exact 
solution of the DMPK equation, presented in Ref. \cite{f95}, by letting
$M \to \infty$ (note that in this case $c_{nk}^{(M)} \to 1$ and Eq. 
(\ref{moments}) remains otherwise unchanged). Another known exact solution is
that obtained in Ref. \cite{z92,mgz94}  for the first two moments of the
conductance of a thick wire. It can be recovered by taking the limits $N,M
\to \infty$, which implies $c_{nk}^{(N)}c_{nk}^{(M)} \to 1$  and the sum is
extended to infinity. Physically, in each of these  limits we are neglecting
(or reducing) the effects of the contact resistances at the lead-sample
interfaces. We remark that the exact expressions for the third moment of
the conductance in both limits (DMPK and thick wire) are presented
here for the first time. 
\par
There are a number of interesting transport
regimes that can be described by our explicit solution. Let us consider here
some of them separately.

\noindent 
(a) The insulating regime ($L \gg \xi$) 
 
In this regime the system becomes an effective disordered 1D wire and all
moments of the conductance decay exponentially with increasing length
as a consequence of wave-function localization. 
Is this case we may evaluate Eq. (\ref{moments}) for $s \gg 1$ using 
the saddle-point method, we find 
 
\begin{eqnarray*} 
\left\langle G/G_0\right\rangle _s &\simeq &4\langle \left( 
G/G_0\right) ^2\rangle _s\simeq \frac{64}9\langle \left( 
G/G_0\right) ^3\rangle _s \\ 
&\simeq &2c_{0,0}^{(N_1)}c_{0,0}^{(N_2)}\left( \frac \pi s\right) 
^{3/2}e^{-s/4}. 
\end{eqnarray*} 
\noindent 
(b) The quasi-ballistic regime ($L_{sys} \gtrsim l$) 
 
In this regime the system is in the transition from boundary surface 
dominated to bulk disorder dominated scattering as its length, $L_{sys}$,
increases from its ballistic value $l$.
We get expressions in this limit by expanding $\left\langle 
(G/G_0)^m\right\rangle _s$ in a power series about $s=0$. We find 
\[ 
\left\langle G/G_0\right\rangle _s =g_c\left( 1-\frac{g_cs}{1-(\alpha g_c)^2}
+\cdots \right)  
\] 
and 
\[ 
{\rm var}(G/G_0)=\frac{(\alpha g_c^2)^2}{1-(\alpha g_c)^2}\left( 1+\frac{%
2(1-4\alpha g_c^2)s}{\alpha g_c\left( 1-(2\alpha g_c)^2\right) }+\cdots 
\right),  
\] 
where $g_c=N_1N_2/(N_1+N_2)$ is the dimensionless contact conductance and
$\alpha \equiv(N_1N_2)^{-1}$. Note that the first term reproduces the 
results of Ref. \cite{b97} for quantum dots. 
  
\noindent  
(c) The semiclassical regime ($N_1,N_2 \gg 1$) 
 
This is the most studied regime of quantum transport and corresponds to 
a situation where quantum interference effects can be treated as a
perturbation to classical transport theory. There are several useful methods
to obtain expressions for  observables in this regime, such as the
diagrammatic impurity average perturbation theory \cite{lsf87}, the large $N$
expansion of the non-linear $\sigma$-model \cite{iwz90}  and the semiclassical
approach using trace formulas \cite{bjs93}. For the present model
we obtain    \[ 
\left\langle G/G_0\right\rangle _s\simeq \frac 1{1/g_c+s} , 
\] 
which corresponds to the expected length dependence of the classical Ohm's
law. For the variance of the conductance we get
\[ 
{\rm var}(G/G_0)\simeq \frac 1{15}\left( 1-\frac{6\gamma _5\gamma 
_1(1+g_cs)-5\gamma _3^2}{\left( 1+g_cs\right) ^6\gamma _1^6}\right),   
\] 
where $\gamma _m\equiv N_1^m+N_2^m$. This result agrees qualitatively with
that of Ref. \cite{iwz90}, where a discrete version of the non-linear
$\sigma$-model has been used. The quantitative difference is not surprising
since the  continuum limit smoothes the ballistic-diffusive crossover (see
also Ref. \cite{mb95}, where a different continuum model has been studied).
In the diffusive limit,
i.e. when $sg_c  \gg 1$, we get simply $\left\langle G/G_0\right\rangle _s
\simeq s^{-1}$ and   ${\rm var}(G/G_0)\simeq 1/15$ in agreement with
rigorous
microscopic diagrammatic  calculations \cite{lsf87}. We remark that one
recovers the result of the ballistic-diffusive crossover predicted by the DMPK
theory simply by taking the limit $N_2 \to \infty$ (or $N_1 \to \infty$) in all
formulas.

\noindent 
(ii) The average density of transmission eigenvalues 
 
We define 
\begin {equation} 
\rho (\tau ,s)=\left\langle {\rm tr}\delta (\tau -tt^{\dagger 
})\right\rangle_s
\label{densdef}  
\end{equation} 
as the average density of eigenvalues of $tt^{\dagger}$ (usually called
transmission eigenvalues).  In our formalism $\rho (\tau ,s)$ can be
obtained from Eq (\ref{central}) through the equation \cite{r96} 
\begin{equation} 
\rho (\tau ,s)=\frac 1{\pi \tau }{\rm Im}\left( \tan (\theta _0/2)\frac \partial {%
\partial \theta _0}\Psi_s (\{\theta \})\right), 
\label{density}  
\end{equation} 
in which $\theta_0$ and $\theta_1$ are related to $\tau$ by
$\cos \theta _0=\cosh  \theta _1=1-2(\tau +i0^{+})^{-1}$. Evaluating Eq.
(\ref{density}) we get  
\begin{eqnarray} 
\rho (\tau ,s) &=&\frac 2{\pi\tau ^2}\sum_{n=0}^{N-1}\int_0^\infty d\mu 
_{nk}\cosh(k\pi)c_{nk}^{(N)}c_{nk}^{(M)}\nonumber  \\  
&&\times P_n\left( 1-2\tau ^{-1}\right)P_{-1/2+ik}(2\tau ^{-1}-1)e^{-\varepsilon _{nk}s}. 
\label{densres} 
\end{eqnarray} 
In the limit $s \to 0$ we recover the result of Ref. \cite{am98} for quantum
dots  \[ 
\rho (\tau ,0)=\tau ^\mu\sum_{n=0}^{N-1}(2n+\mu+1)\left\{ P_n^{(\mu,0)}(1-2\tau 
)\right\} ^2 ,
\]
where $\mu=M-N$ and $P_n^{(\alpha,\beta)}(x)$ is the Jacobi polynomial. 
A useful application of Eq. (\ref{densres}) is the calculation of the ensemble 
average of arbitrary linear statistics, i.e. observables that can be written
in the form $A={\rm tr}\left[ f_A(tt^{\dagger })\right]$. Using Eq.
(\ref{densdef}) one can see that  
\begin{equation}  
\left\langle
A\right\rangle _s=\int_0^1\rho (\tau ,s)f_A(\tau ).  
\label{formula} 
\end{equation} 
As a particular application of this formula we consider the shot-noise power,
which  is given by $P=P_0{\rm tr}\left[ tt^{\dagger }(1-tt^{\dagger
})\right]$, where $P_0=2e|V|G_0$ ($V$ is the applied voltage), and thus
$f_P(\tau)=\tau(1-\tau)$. Using Eq. (\ref{formula}) we find  
\[  \left\langle
P/P_0\right\rangle _s=2\sum_{n=0}^{N-1}\int_0^\infty d\mu 
_{nk}p_{nk}c_{nk}^{(N)}c_{nk}^{(M)}e^{-\varepsilon _{nk}s} , 
\] 
where $p_{nk}=3/4-k^2+n(n+1)$. 
 
We shall now sketch the proof of Eq. (\ref{central}). 
The starting point is the supersymmetric non-linear $\sigma$-model
representation of  the ensemble average of the generating function (Eq.
(\ref{genfun})), which reads  \cite{bf96}
\[ 
\Psi_s (\{\theta \})=\int dQ^{\prime }\int dQ^{\prime \prime 
}V_2(Q_0,Q^{\prime })W_s(Q^{\prime },Q^{\prime \prime })V_1(Q^{\prime \prime 
},Q) 
\] 
where $V_i(Q,Q^{\prime })={\rm Sdet}^{-N_i}(Q+Q^{\prime })$, $i=1,2$, (Sdet
stands for the superdeterminant in the convention of Ref. \cite{vwz85}) and
$dQ$  in the invariant measure of the coset space ${\cal
C}=U(1,1/2)/(U(1/1)\otimes U(1/1))$.  

This non-linear $\sigma$-model representation  can be
rigorously derived  from a number of sofisticated stochastic approaches, such
as the IWZ model \cite{iwz90} or the gaussian white noise potential
\cite{e83,e97}. Here we take, however, the point of view of regarding
the non-linear $\sigma$-model as an independent minimal field theoretical
description of the system, in the sense that it is  free from all the
statistically irrelevant details of specific models (either microscopic or
random-matrix approaches). The  price one has to pay for this model
independence is that the localization length becomes a free parameter and the
theory merely stablishes a scaling law for the physical observables.

It has been shown \cite{e83} that points in ${\cal C}$ can be represented by
supermatrices of the form   
\[  
Q=U\left(  
\begin{array}{cc} 
\cos \hat{\theta} & i\sin \hat{\theta} \\  
-i\sin \hat{\theta} & -\cos \hat{\theta} 
\end{array} 
\right) U^{-1}\,\,;\,\,\,\,\,\,\,\,\,\,\,U=\left(  
\begin{array}{cc} 
u_1 & 0 \\  
0 & u_2 
\end{array} 
\right),  
\] 
where $\hat{\theta}\equiv {\rm diag}(i\theta _1,\theta _0)$; $\theta_1
>0$, $0 < \theta_0 < \pi$, and $u_1$, $u_2$  are $2\times 2$ unitary
supermatrices. In particular the origin of ${\cal C}$ is represented by
$Q_0={\rm diag}(1,1,-1,-1)$. The function $W_s(Q^{\prime },Q^{\prime \prime
})$ is the diffusion kernel of ${\cal C}$ and can be written as a path
integral  given by 
\[   
W_s(Q^{\prime },Q^{\prime \prime })=\int
\prod_{x=0}^sdQ(x) \exp \left( {1 \over 16}
\int_0^s dx {\rm Str}(\partial_x Q)^2 \right), 
\] 
with boundary conditions $Q(0)=Q^{\prime \prime }$ and $Q(s)=Q^{\prime }$.
Here Str is the supertrace as defined in Ref \cite{vwz85}. As shown in Ref.
\cite{bf96}, the high degree of
symmetry in this approach implies that $\Psi_s (\{\theta \})$ satisfies the
following diffusion equation  
\begin{equation}  
\left( \partial _s-\Delta _{\hat{\theta}}\right) \Psi_s
(\{\theta\})=0  \label{diffeq}  
\end{equation}  
with initial condition  
\[ 
\Psi_0 (\{\theta \})=\int dQ^{\prime }V_2(Q_0,Q^{\prime 
})V_1(Q^{\prime },Q), 
\] 
where $\Delta _{\hat{\theta}}$ is the radial part of the Laplace-Beltrami 
operator of ${\cal C}$. The initial condition integral has been studied in
Ref. \cite{am98} and can be expressed in the following rather convenient form 
\begin{eqnarray*}  
\Psi_0 (\{\theta \})&=&1+(\cos \theta _0-\cosh \theta
_1)\sum_{l=0}^{N-1}{(-1)^{l+1} \over 2}\\ 
&&\times {\sin ^{2l}(\theta _0/2) \over \sinh ^{2l+1}(\theta _1/2)}
\left[ 
f_{N-l-1}^{(\mu)}(\theta _0)g_{N-l-1}^{(\mu)}(\theta  _1)-1
\right] 
\end{eqnarray*} 
where $f_n^{(\mu)}(\theta _0)=F(-n,-n-\mu;-2n-\mu;\sin ^2(\theta _0/2))$ and 
$g_n^{(\mu)}(\theta _1)=F(n+1,n+\mu+1;2n+\mu+2;-\sinh ^2(\theta _1/2))$ are
hypergeometric functions. We solve Eq. (\ref{diffeq}) by expanding $\Psi_s
(\{\theta\})$ in a complete set of eigenfunctions of $\Delta _{\hat{\theta}}$
(for a detailed discussion of this procedure see Ref \cite{mgz94}). The final
result is Eq. (\ref{central}).
\par
In summary, we have studied a unified field-theoretical description of the
quantum dot - quantum wire crossover for system with broken time-reversal
symmetry. Using a generating function approach we have found closed
expressions for a number of important transport properties, such as the first
three moments of the conductance, the average shot-noise power and the
average density of transmission eigenvalues. Extensions of our results to 
other symmetry classes (orthogonal and symplectic) and to more elaborate
correlation functions (with explicit dependence on energy and/or magnetic
field) appear to be possible by means of the theory of superharmonic analysis
presented in Ref. \cite{mgz94}. An interesting immediate application of
the techniques presented here would be the derivation of extensions of
the variance \cite{cmbr93} and the length-correlation \cite{m97} formulas to
all physical regimes and  length scales.

This work was partially supported by  CNPq, FACEPE and FINEP (Brazilian
Agencies).

\end{document}